\documentstyle[aps,graphicx,multicol]{revtex}
\def\bk{{\bf k}}
\def\bK{{\bf K}}
\def\br{{\bf r}}

\def\bp{{\bf p}}

\def\bQ{{\bf Q}}
\def\bv{{\bf v}}
\def\cH{{\mathcal H}}
\def\vep{\varepsilon}

\def\ad{\alpha_D}
\def\vd{v_{\Delta}}
\def\rb{\right}
\def\lb{\left}
     
\def\im{i}

\begin{document}

\renewcommand{\baselinestretch}{0.5}

\title{Dirac quasiparticles in the mixed state}
\author{Daniel Knapp, Catherine Kallin and A.\ J.\  Berlinsky}
\address{Department of Physics and Astronomy, McMaster University, Hamilton, Ontario, Canada L8S 4M1}
\date{\today}

\maketitle

\begin{abstract}
Energies and wave functions are calculated for $d$-wave quasiparticles
in the mixed state using the formalism of Franz and Te\u{s}anovi\'{c}
for the low-lying energy levels. The accuracy of the plane-wave
expansion is explored by comparing approximate to exact results for a
simplified one-dimensional problem, and the convergence of the plane
wave expansion to the two-dimensional case is studied.  The results
are used to calculate the low-energy tunneling density of states and
the low-temperature specific heat, and these theoretical results are
compared to semiclassical treatments and to the available data.
Implications for the muon spin resonance measurements of vortex core
size are also discussed.
\end{abstract}


\begin{multicols}{2}

\section{Introduction}
The nature of the low-lying excitations in the mixed state of a $d$-wave 
superconductor is both an interesting quantum mechanics problem and 
important for understanding the behavior of high temperature superconductors
 in a magnetic field\cite{volovik1,gorkov,anderson,franz1,janko}.  
Volovik\cite{volovik1} first studied 
this problem in the semiclassical limit, where the $d$-wave quasiparticles 
are Doppler shifted by the local superfluid density.  The shifting of
quasiparticle energies results in a non-zero density of states at zero
energy proportional to the square root of the magnetic field.  
Volovik's solution has been applied to calculations of the specific 
heat\cite{kam,geneva,kubert} thermal conductivity\cite{franz1b,hirsch}, and 
nuclear magnetic relaxation rates\cite{wortis1,wortis2}.  It has 
motivated useful discussions of the scaling behavior of the specific heat 
by Simon and Lee\cite{simonlee}, Kopnin and Volovik\cite{volovik2}, and others. 

The presence of a magnetic field and its associated vortex lattice affects the 
motion of quasiparticles in four distinct ways.  First, the quasiparticles, 
which carry current, move in the magnetic field which is approximately uniform 
for an extreme type-II superconductor.  Second, although the field is 
approximately uniform, it is not exactly so, and therefore the quasiparticles 
experience magnetic field gradients. However the direct effect of these 
gradients is rather small. Third, there are supercurrents associated with the 
curl of the field, and the quasiparticle energies are affected by the 
corresponding superfluid velocity through which they move.  For a uniform 
superfluid velocity field, the effect would be a simple Doppler shift of the 
energies. However, for inhomogeneous superfluid velocities the problem is more 
complicated. Fourth and finally, the magnitude of the superconducting order 
parameter is inhomogeneous in the mixed state, although this is mainly apparent
within a coherence length of each vortex core where  this magnitude falls to 
zero. For an extreme type-II superconductor, this represents a very small 
fraction of the sample for fields well below $H_{c2}$.

Volovik's approach neglects the magnetic field and its gradients as well as the inhomogeneous order parameter amplitude and focuses only on the local supercurrent velocity. It assumes that the quasiparticle wave function can be thought of as a wave packet which is localized in a region over which the magnitude and direction of the superfluid velocity are relatively uniform. The energy of a low-lying, $d$-wave quasiparticle depends linearly on $\vec  q=\vec k-\vec k_\alpha$, where $\vec k_\alpha$ is the wave vector of the nearest node. If a quasiparticle is localized in a region of size smaller than $1/|\vec  q|$, then the spread in its energy will be larger than its average energy and the wavepacket picture does not work.  For the superfluid velocity to be relatively uniform in a region, the size of the region must be smaller than the distance to the nearest vortex core and certainly smaller than, $d$, the distance between vortices.  Let us apply the above considerations to the lowest energy quasiparticle band.  This corresponds to a quasiparticle, near node $\nu$, with wave vector $\vec q$ perpendicular to $\vec k_{\nu}$ localized in a region of size $\ell$.  For the wavepacket picture to apply, the energy, $E$, of the quasiparticle must be greater than $\hbar v_\Delta\pi/\ell$, where $v_\Delta$ is the quasiparticle velocity along $\vec q$.  However for the superfluid velocity to be uniform in the region of size $\ell$ it is necessary that $\ell\ll d$. Combining these two conditions we obtain the requirement that $E \gg \hbar v_\Delta\pi/d$. For energies less than this the wavepacket picture breaks down and a full quantum mechanical picture is needed. This energy range is readily accessible via specific heat measurements below $10$\,K in fields of one to several Tesla. It is this energy region that is the main focus of this paper.

Recently, Franz and Te\u{s}anovi\'{c}\cite{franz1} (FT) have derived a 
quantum mechanical theory of the mixed state of a $d$-wave superconductor, 
which involves a singular gauge transformation that maps the original problem of 
superconducting quasiparticles in a magnetic field onto an equivalent one of 
quasiparticles in a periodic potential.  The latter problem may be solved using
conventional band-structure methods.

In this paper, we investigate the low-energy properties of a $d$-wave 
superconductor in the mixed state using the theory derived by Franz and 
Te\u{s}anovi\'{c}\cite{franz1}.  The most direct experimental probes of these 
properties are the low-energy tunneling density of states and the low 
temperature specific heat.  In order to calculate these quantities reliably, 
we have investigated the numerical problem of Dirac quasiparticles in the 
periodic potential of the vortex lattice, focusing on the simplifications that 
result from the fact that the anisotropy of the Dirac cones, $\ad=v_F/v_\Delta 
\gg 1$. As discussed by Mel'nikov\cite{melnikov1}, such large anisotropy makes 
the problem approximately one-dimensional.  Mel'nikov described how to obtain 
solutions to the one-dimensional problem, but he then confined his analysis to
the semiclassical versions of these solutions.  We have explicitly evaluated the
quantum mechanical solutions in this one-dimensional limit and used them as a 
test of the accuracy of approximate plane-wave solutions.  We then show how to 
improve on the one-dimensional solutions by including a small number of plane 
wave basis states for the transverse direction, and we study the convergence of
this approach.

The remainder of this paper is organized as follows. Section II addresses the 
computational problem of calculating quasiparticle energies in the lowest bands
in the magnetic Brillouin zone, comparing exact and plane-wave-expansion 
solutions for the simplified one-dimensional problem and then comparing 1-D 
and 2-D plane-wave expansion solutions for various choices of plane-waves.  
Section III presents results for the local tunneling density of states, Sec.\
IV re-interprets recent muon spin resonance measurements of
the vortex core size in terms of a scaling picture, and Sec.\ V 
presents calculations of the low temperature specific heat, comparing 
them to predictions based on Volovik's approach and to experimental data.

\section{The Computational Problem:  calculating the energies in the lowest 
bands in the magnetic Brillouin zone}

The quasiparticle wavefunctions are described by the BdG equations, $\cH\psi=\epsilon\psi$ where $\psi=(u(\br), v(\br))^T$, and 
\begin{equation} \label{eqn:bdgH}
\cH=\left( \begin{array}{cc}
\hat{\cH_e}&\hat{\Delta}\\
\hat{\Delta}^*&-\hat{\cH_e^*}
\end{array} \right),
\end{equation}
with $\cH_e=(\bp-\frac{e}{c}{\bf A})^2-\epsilon_F$.  The gauge invariant form of the gap operator, $\hat{\Delta}$, for a $d$-wave superconductor can be written as (see Ref. \onlinecite{franz2} for the details)
\begin{equation}
\hat{\Delta}=\frac{1}{p_F^2}\{\hat{p_x}, \{\hat{p_y}, \Delta(\br)\}\}-\frac{\im}{4}\frac{1}{p_F^2}\Delta(\br)(\partial_x\partial_y \varphi),
\end{equation}
where, for notational convenience, we have chosen to orient our axes along the directions of the gap nodes, at an angle of $\pi/4$ with respect to the 
orientation of the CuO$_2$\ planes.  $p_F$ is the Fermi momentum, and 
$\Delta(\br)=|\Delta(\br)|e^{\im\varphi(\br)}$ is the Ginzburg-Landau order 
parameter.  Since we are working in the intermediate field regime 
($H_{c1}\alt B<<H_{c2}$) of an extreme type II superconductor, we can assume 
that the magnitude of the gap is constant everywhere, except at the vortex 
cores, and that the magnetic field distribution and local superfluid velocity 
can be described by the London model\cite{london}. 

In order to diagonalize 
the Hamiltonian in Eq.\ (\ref{eqn:bdgH}) one would like to remove the order 
parameter phase from the off-diagonal components of $\cH$.  It is desirable 
to choose a transformation which is both single-valued and treats the 
particles and holes on an equal footing. 
This is accomplished by the bipartite, singular gauge transformation of 
FT:\cite{franz1}
\begin{equation} \label{eqn:gaugetrans}
\cH \to U^{-1}\cH U,~~~~U=\left(\begin{array}{cc}e^{\im\varphi_A(\br)}&0\\0&e^{-\im\varphi_B(\br)}\end{array}\right),
\end{equation}
where $\varphi_A(\br)+\varphi_B(\br)=\varphi(\br)$, and $\varphi_A(\br)$ and $\varphi_B(\br)$ are the contributions to the order parameter phase from the $A$ and $B$ sublattices of the vortex lattice.  The sublattices are chosen so that there are equal numbers of $A$ and $B$ vortices, with two vortices per magnetic unit cell of the vortex lattice.  The vortex lattice configuration analyzed in this paper is shown in Fig.\ \ref{fig:lattice}.  Note that the fact that the $x$- and $y$-axes of the $A$ and $B$ sublattice
unit cells are oriented along nodal directions means that nearest-neighbor lines of
vortices are oriented along the $a$- and $b$-axes of the underlying crystal lattice.

Under this transformation $\cH$ becomes
\begin{equation} \label{eqn:abH}
\cH_{\text{AB}}=\left(\begin{array}{cc}\frac{1}{2 m}(\bp+m\bv_s^A)^2-\epsilon_F&\hat{D}\\
\hat{D}&-\frac{1}{2 m}(\bp-m\bv_s^B)^2+\epsilon_F \end{array}
 \right),
\end{equation}
where
\begin{equation}
\hat{D}=\frac{\Delta_0}{p_F^2}\left[\hat{p}_x+\frac{m}{2}(v^A_{sx}-v^B_{sx})\right] \times \left[\hat{p}_y+\frac{m}{2}(v^A_{sy}-v^B_{sy})\right],
\end{equation}
with the superfluid velocities 
\begin{equation} \label{eqn:vs_mu}
\bv^{\mu}_s(\br)=\frac{1}{m}\left(\hbar \nabla \varphi_{\mu}-\frac{e}{c}{\bf A}\right),~~~~~\mu=A,B.
\end{equation}
Note that $\bv^A_s(\br)+\bv^B_s(\br)=2 \bv_s(\br)$.  Since the vortex lattice is periodic, the superfluid velocities can be written as Fourier sums  
\begin{equation} \label{eqn:fourierv}
\bv_s^{\mu}(\br)=\frac{2\pi\hbar}{md^2}\sum_{\bK\neq0}\frac{\im\bK\times\hat{z}}{K^2}e^{\im\bK\cdot(\br+\vec{\delta}^{\mu})},           
\end{equation}
where $\bK=(2\pi/d)(m_x,m_y)$, $d=\sqrt{2\Phi_0/B}$ is the size of the magnetic unit cell, and $\vec{\delta}^{\mu}=\pm(d/4,d/4)$ is the displacement of $A$ or $B$ vortices from the center of the unit cell (see Fig.\ \ref{fig:lattice}).

Linearizing the Hamiltonian in Eq.\ (\ref{eqn:abH}) at the node $\vec k = (k_F,0)$ we find that $\cH_{\text{AB}}\simeq \cH_0+\cH'$ with 
\begin{equation} 
\label{eqn:H0}
{\mathcal H}_0=\left(
\begin{array}{cc}
v_F p_x&v_{\Delta} p_y\\
v_{\Delta} p_y&-v_F p_x\\
\end{array}\right)
\end{equation}
and
\begin{equation} \label{eqn:H'}
\cH'=m \left(
\begin{array}{cc}
v_F v_{sx}^A & \frac{1}{2}v_{\Delta}(v_{sy}^A-v_{sy}^B)\\
\frac{1}{2}v_{\Delta}(v_{sy}^A-v_{sy}^B) & v_F v_{sx}^B
\end{array}
\right),
\end{equation}
where $v_F$ is the Fermi velocity, and $v_{\Delta}=\Delta_0/p_F$ is the slope of the gap at the node.  

At the node $\vec k=(k_F,0)$ the free Dirac Hamiltonian $\cH_0$ has the familiar Dirac cone spectrum 
\begin{equation} 
\vep \simeq \pm \hbar v_F \sqrt{k_x^2+k_y^2/\ad^2},
\end{equation}
where $(k_x, k_y)=0$ at the node.  The quasiparticle momentum along the nodal direction is $k_x \sim \vep/\hbar v_F$ with a corresponding wavelength of $\lambda_x \sim \hbar v_F/\vep$.  If the energy is low enough ($\vep\alt\hbar v_F/r_x$), $k_x$ will be confined to the first magnetic Brillouin zone (see Fig.\ {\ref{fig:mbz}) and the wavelength $\lambda_x$ will exceed the intervortex distance $r_x$, crossing the boundaries of several unit cells of the vortex lattice. For large values of the anisotropy $\ad$ the momentum parallel to the Fermi surface $k_y \sim \ad k_x$ will be much larger than that along the nodal direction.  The quasiparticle wavefunction will thus be localized in the direction parallel to the Fermi surface, but will be extended and will feel the average effect of the superfluid velocity fields of several vortices along the nodal direction.

Since the potential ${\mathcal H}'$ is periodic we can expand the quasiparticle wavefunction in the plane-wave basis:
\begin{equation} \label{eqn:bloch2d}
\Psi_{\bf k}(\br)=\sum_{\bK}e^{\im(\bf k+\bK)\cdot\br}\psi_{\bK}(\bf k).
\end{equation}
The periodic potential of the vortex lattice will be responsible for the interaction of the Fourier components $\psi_\bK(\bk)$ and $\psi_{\bK^\prime}(\bk)$.  If we are interested only in energies below a cutoff energy, $E_c$, for which the momentum $k_x$ 
lies well within the first MBZ we can make the approximation that the quasiparticle wavefunction is one-dimensional and ignore the interaction of those Fourier components that are at different values of $K_x$. We therefore write
\begin{equation} \label{eqn:bloch1d}
\Psi_{\bk}(\br)\simeq e^{\im k_x x}\sum_{K_y}e^{\im(k_y+K_y) y}\psi_{K_y}(\bk).
\end{equation}
If, however, $E_c$ is high enough that $k_x$ exceeds the boundaries of the first MBZ (see Fig.\ \ref{fig:mbz}) 
we can make the assumption--since the Fourier sum is dominated by components whose values of $\bK$ are bounded by the constant energy contour at $E_c$--that
\begin{equation} \label{eqn:blochq1d}
\Psi_{\bk}(\br)\simeq\sum_{K_y}^{K_c}\sum_{K_x}^{K_c/\ad}e^{\im(\bk+\bK)\cdot\br}\psi_{\bK}(\bk),
\end{equation}
where $K_c$ is the cutoff wave vector along the {\emph{y}} direction.

Such plane-wave expansions can be computed numerically to obtain the excitation 
spectrum for the quasiparticles in a periodic vortex lattice. The solution to the 
problem using Eq.\ (\ref{eqn:bloch2d}) has been studied in detail by 
FT\cite{franz1}, whereas
 Marinelli, Halperin and Simon (MHS) \cite{marinelli} studied solutions to ${\mathcal H}_{AB}$
defined in Eqs.\ (\ref{eqn:H0}) and (\ref{eqn:H'}) in position space.  Both
groups found that the convergence of the plane-wave expansions was slow.
Since we are specifically interested in the low energy and low temperature properties which are largely determined by the lowest band of the excitation spectrum, we will focus next on obtaining an analytical solution to the linearized Hamiltonian with the
 approximation that, for large $\ad$, the quasiparticle wavefunctions are approximately one-dimensional.  Having obtained both analytical and numerical solutions to this one-dimensional problem, we will then examine how adding more transverse wave vectors
, as in Eq.\ (\ref{eqn:blochq1d}), allows us to approach the exact numerical 
two-dimensional result, using
Eq.\ (\ref{eqn:bloch2d}).

\subsection{The 1-D analytical solution}

At low energies and for large values of $\ad$ the wavefunctions are localized in the 
{\emph{y}} direction and extended along the {\emph{x}} direction.\cite{melnikov1}  This suggests the following basis as a useful starting point:
\begin{equation}
\Psi_{\bk}(\br)=\sum_{K_x}e^{\im(k_x+K_x)x}\psi(k_x+K_x,y).
\end{equation}
As we shall see, the Fourier components  $\psi(k_x+K_x,y)$, for different $K_x$, are spatially well separated in the {\emph{y}} direction.  Their interaction is consequently negligible, and we can assume that the Hamiltonian is diagonal in the quantum number $K
_x$.  This allows us to replace the periodic potential ${\mathcal H'}$ in Eq.\ (\ref{eqn:H'}), which in principle scatters quasiparticles between states with different values of $K_x$, with its effective potential averaged in the {\emph{x}} direction which is 
diagonal in $K_x$.  The result is
\begin{equation} \label{eqn:1dH}
{\mathcal H}_{\text{1D}}=\left( 
\begin{array}{cc}
q_x^n   &  -\frac{\im}{2 \pi \ad} \frac{d}{dz} \\
 -\frac{\im}{2 \pi \ad} \frac{d}{dz} & -q_x^n
\end{array}
\right)
+
\left( 
\begin{array}{cc}
\Phi(z-\frac{1}{4})  &  0 \\
0  &  \Phi(z+\frac{1}{4})
\end{array}
\right),
\end{equation}
where $\mathcal{H} \psi = \epsilon \psi$, $\epsilon=E/E_0$ where 
$E_0=2 \pi \hbar v_F/d$, $z=y/d$, and $\Phi(z)=z-(n+\frac{1}{2})$ 
where $n<z<n+1$. Note that in these units $n-\frac{1}{2}<q_x^n<n+\frac{1}{2}$, 
where $q_x^n=(d/2\pi)(k_x+K_x)$.  The potential along the {\emph{y}} direction 
consists of two periodic sawtooth functions with discontinuities lying along 
the averaged vortex lines of the $A$ and $B$ sublattices. At sufficiently low 
energies the quasiparticles will be bound in the {\emph{y}} direction by the potential
barriers which lie at the discontinuities in $\Phi(z)$. Our picture is thus one
of quasiparticles that travel as plane-waves along the nodal direction but are 
bound within potential wells--created by the averaged vortex lattice--in the direction parallel to the Fermi surface.  Note that, at low energies, the Fourier components $\psi(k_x, y)$ and $\psi(k_x+K_x, y)$ have negligible overlap, as they  lie in \textit{separate} potential wells along the {\emph{y}} direction. 
 
By making the substitution 
\begin{displaymath}
\varphi(z)=\frac{1}{2}(\hat{\sigma}_x+\hat{\sigma}_z)\psi(z),
\end{displaymath}
Eq.\ \ref{eqn:1dH} can be rewritten as
\begin{equation} \label{eqn:phi}
\left\{\hat{\sigma}_z \left( -\frac{\im}{a}\frac{d}{dz}\right)+\frac{\Phi_1(z)}{2}-\epsilon+\hat{\sigma}_x\left(q_x^n+\frac{\Phi_2(z)}{2}\right)\right\}\varphi(z)=0\,,
\end{equation}
where $a=2\pi\ad$ and $\hat{\sigma}_i$ are the Pauli matrices.  The function \[ \Phi_1(z)=\Phi(z-\frac{1}{4})+\Phi(z+\frac{1}{4})\] is a sawtooth with a slope of $+2$ and a period of $1/2$, and the function \[\Phi_2(z)=\Phi(z-\frac{1}{4})-\Phi(z+\frac{1}{4})\] is a step function which oscillates between $-1$ and $+1$ with a period of $1$. Since the potential is periodic we can solve Eq.\ (\ref{eqn:phi}) within a unit cell and use Bloch's theorem to extend the solution over all of $z$.  The solution within a unit cell (see Appendix \ref{1dappendix} for the details) is given in terms of the parabolic cylinder functions $D_p(z)$:\cite{g&r}
\begin{equation} \label{eqn:solution1}
\varphi_1(z)=\left(
\begin{array}{c}
D_{(\im a/2)(q_x^n+\frac{1}{4})^2}\left[\pm\sqrt{2\im}\,z_1 \right] \\ \\
\mp \sqrt{\frac{\im a}{2}}(q_x^n+\frac{1}{4}) D_{(\im a/2)(q_x^n+\frac{1}{4})^2-1}\left[\pm\sqrt{2\im}\, z_1 \right]
\end{array}
\right)
\end{equation}
for $n-\frac{1}{4}<z<n+\frac{1}{4}$ and
\begin{equation} \label{eqn:solution2}
\varphi_2(z)=\left(
\begin{array}{c}
D_{(\im a/2)(q_x^n-\frac{1}{4})^2}\left[ \pm\sqrt{2\im}\,z_2 \right] \\ \\
\mp \sqrt{\frac{\im a}{2}}(q_x^n-\frac{1}{4})  
D_{(\im a/2)(q_x^n-\frac{1}{4})^2-1}\left[ \pm\sqrt{2\im}\,z_2 \right]
\end{array}
\right)
\end{equation}
for $n+\frac{1}{4}<z<n+\frac{3}{4}$, with $z_1=\sqrt{a}(z-\epsilon-n)$ and $z_2=\sqrt{a}(z-\epsilon-n-\frac{1}{2})$.

These solutions can be matched at the boundaries of the unit cell (see  Appendix B) 
to obtain an exact excitation spectrum for the one-dimensional, averaged Hamiltonian.
The resulting spectrum for anisotropy $\ad=7$ is shown in Fig.\ \ref{fig:anal7}.  
It is useful to note that the energy scale $E_0$ is approximately given by $E_0\approx 185\sqrt{B}\;\text{T}^{-1/2}\,\text{K}$.

\subsection{Comparison of 1-D analytical and plane-wave expansion results} 
\label{sec:sub1d}

Using the plane-wave expansion of Eq.\ (\ref{eqn:bloch1d}) we can numerically 
diagonalize the Hamiltonian to obtain an excitation spectrum that can be compared 
with the analytical results.  To the numerical accuracy of the diagonalization,
these two methods yield identical results for the dispersion along $k_x$ as 
shown in Fig.\ \ref{fig:anal7}, where 61 reciprocal lattice vectors (RLVs) 
have been kept in the plane-wave expansion.\cite{convergence}  The dispersion
along $k_y$ calculated from the 1-D plane-wave expansion
is also shown in Fig.\ \ref{fig:anal7}. As discussed by MHS,
\cite{marinelli} the dispersion away from the $\Gamma$ point along $k_y$
is more strongly renormalized by the supercurrents, leading to an enhanced 
effective $\ad$.  For
$\ad>10$ there is essentially no discernable dispersion along $k_y$ for the
lowest bands (as calculated in either the 1-D or full 2-D plane-wave 
expansions), 
further suggesting the validity of a one-dimensional approximation.

Since both the energy and momentum axes scale as $\sqrt{B}$ these spectra 
apply to all values of the magnetic field within $H_{c1}<<B<<H_{c2}$.  As 
the anisotropy increases, the gap between the lowest band of the spectrum 
and the $E=0$ axis quickly narrows.  At $\ad=14$ (Fig.\ \ref{fig:1d2d14}) 
the spectrum is close to 
forming a line quasinode, in agreement with the results of FT\cite{franz1},
and of MHS \cite{marinelli}, which suggest that a line 
quasinode first appears at $\ad\simeq15$. 

\subsection{Comparison of 1-D and approximate 2-D plane-wave calculations}
\label{sec:sub2d}

The results of numerical diagonalization calculations of the excitation spectrum of the quasiparticles in the 1-D averaged potential and the exact 2-D potential at different values of the anisotropy are shown in Figs.\ \ref{fig:1d2d14} and \ref{fig:1d2d20}. The 1-D spectra show good qualitative agreement with the 2-D spectra, capturing the major features of the lowest bands, including the line quasinodes that form at large values of $\ad$.  However, the 1-D treatment is unable to accurately represent, \textit{quantitatively}, the behavior of the full 2-D spectrum. In particular, as can be seen from Figs.\ \ref{fig:1d2d14} and \ref{fig:1d2d20}, the 1-D approximation cannot be used to quantitatively determine the size of the minigaps which lie along the line quasinodes.   An analysis of the 1-D spectra for several values of $\ad$ shows that the size of the 
smallest minigap at fixed $\ad$ is $\delta_g \propto e^{-m\ad}$\ where 
$m\approx 0.18$.
Unfortunately, the slow convergence of the 2-D reciprocal lattice sums--due 
to the divergence of the superfluid velocity at the vortex cores (discussed in 
more detail by Vafek \textit{et al.}\cite{franz2})--makes it very difficult 
to accurately determine the size of these minigaps in the full 2-D 
calculation.

Nonetheless, we believe that the 1-D treatment (which is far less 
computationally intensive than the 2-D problem) captures and elucidates 
the important physics of the lowest bands of the quasiparticle excitation 
spectrum and is therefore a useful tool that helps us understand the 
physical behavior of the quasiparticles in the mixed state.   In particular
we will use the 1-D energies and wave functions to calculate the
local tunneling density of states and the specific heat.

Next we compare the results of the 1-D calculations to finite 2-D plane
wave expansions using a grid of $N_x\times N_y$ reciprocal lattice
vectors.  For example, in Fig. \ref{fig:conv14} results are shown for $\ad=14$,
comparing the 1-D case, $N_x=1,\ N_y=41$, to $N_x=$ 5, 9, 13, 21,
 and 29,\
$N_y=41$.  Similar results are shown in Fig. \ref{fig:conv20}
for $\ad=20$ and
$N_y=61$.  One of the most striking features of both figures is the
complete insensitivity of the linear branch near the $\Gamma$ point to the
number of plane-waves in the calculation.  For this branch, it appears
that the 1-D energies are essentially exact. For other low-energy branches
and general points in the Brillouin zone, the plane-wave expansion seems
to converge smoothly. The only pathological behavior occurs near the
quasinodes, where both the positions and the values of the minima
converge slowly.

\section{Local Tunneling Density Of States}
In this section we show results of calculations of the local tunneling density of states (TDOS) of the quasiparticles in the lowest band of the energy spectrum using the one-dimensional plane-wave expansion of Eq.\ (\ref{eqn:bloch1d}).  The TDOS is:\cite{gygi,wang}
\begin{eqnarray}\label{eqn:ldos}
N({\bf r},E)&=&-\frac{2}{N_k}\sum_{{\bk},\mu,\nu}\ \ |u_{{\bk},\mu,\nu}({\bf r})|^2f'(\vep_{\bk,\mu}-E)\ \nonumber \\
&+&\ |v_{{\bk},\mu,\nu}({\bf r})|^2f'(\vep_{\bk,\mu}+E), 
\end{eqnarray}
where $f'(x)$ is the derivative of the Fermi function, $\bk$ is the set of wave vectors in the magnetic Brillouin zone, $\mu$ is the set of energy bands (restricted in this case to the lowest positive and negative energy bands) and $\nu$ is the set of four Dirac nodes. The normalization factor $2/N_k$ is equal to the number of spins divided by the number of wave vectors in the magnetic
Brillouin zone.  The extra factor of 2 is an artifact of the normalization in the diagonalization routine used.

The plane-wave expansion was done at the node $\vec k=(k_F, 0)$.  It is easy to show that by taking $\vep_{{\bk},\mu} \to -\vep_{{\bk},\mu}$ in Eq.\ (\ref{eqn:ldos}) one obtains the contribution from the opposite node at $\vec k=(-k_F,0)$.  Within the 1-D approximation, these two nodes give the {\emph{y}} dependence of the TDOS, and the other two nodes at $\vec k=(0, \pm k_F)$ give the {\emph{x}} dependence.  The {\emph{y}} (or {\emph{x}}) dependence of the TDOS at 1 Kelvin and at a field of 1 Tesla, for two different values of the anisotropy $\ad$, and at three different energies is shown in Figs.\ \ref{fig:ldos14} and \ref{fig:ldos20}. 

One can see that the TDOS has the periodicity of, and is sharply peaked at, the vortex lines.  The TDOS falls to a broad minimum in the regions between the vortices.  The shoulders on either side of the peaks come from the states within the lines of quasinodes that form at large values of $\ad$. At $\ad=20$, the size of the gap in the line quasinode has decreased and a second line quasinode has started to appear (see Fig.\ \ref{fig:1d2d20}).  Both these features contribute to the very distinct shoulders on either side of the peak in the TDOS in Fig.\ \ref{fig:ldos20}. 

Figure \ref{fig:ldos2d} shows the zero-bias two-dimensional TDOS as a sum over the four nodes. 
This result is in qualitative agreement with the semiclassical calculation of the TDOS by Mel'nikov \cite{melnikov2}.  The vortex lattice geometry of our paper is, in Mel'nikov's notation, a Type II lattice with $\sigma=1/2$. This gives a TDOS that is proportional to 
\begin{equation} 
F_1=\lb|\Phi\lb(\frac{x}{d/2}\rb)\rb|+\lb|\Phi\lb(\frac{y}{d/2}\rb)\rb|,
\end{equation}
where $\Phi(z)=2z-(2m+1)$.  The semiclassical TDOS of Mel'nikov thus has the profile of a triangle wave along the {\emph{x}} and {\emph{y}} directions.  The fully quantum mechanical results shown here follow this profile, but exhibit additional structure that arises from the quasiparticle states near the quasinodes.

We note that only half of the bright spots in Fig. \ref{fig:ldos2d}
lies at vortex positions
while the other half lies halfway between vortices.  For example, in
Fig. \ref{fig:ldos2d} the bright 
spots at the corners and at the center of the figure
might correspond to vortex sites. The other bright spots are then the
result of the overlap of the sharply peaked tunneling density of states
which extends from each vortex, parallel to the four node directions. It
is an artifact of the 1-D model that, for the case of a square lattice,
these overlaps have a peak tunneling density of states equal to that of a
vortex core.  This artifact is less evident in more general, centered
rectangular lattices or, in particular, for the hexagonal 
lattice.\cite{melnikov1,melnikov2}

\section{Muon Spin Resonance}

Two important simplifying assumptions in this model are that the
superconducting coherence length is negligible and that the penetration depth is large compared to the distance between vortices.  As a consequence of these assumptions,
the intervortex spacing is the only length scale in the problem.  This
allows us to present results scaled to this length as is done above
for the tunneling density of states.

In addition to the tunneling density of states, one could also use the 
wave functions generated by these calculations to compute the 
pattern of the two-dimensional supercurrent density.  This would, of
course, not be a self-consistent result, but it would be an improvement
over the initial form for the supercurrent density corresponding to 
Eq. (7).  Without actually doing this calculation, we know that
the resulting pattern would be a function of ${\bf r}/d$ and hence that all
lengths would scale as $1/\sqrt{B}$. 

This picture, in which the vortex lattice constant provides the only
length scale, is supported by the self-consistent calculations of
Franz and Tesanovic for a single $d$-wave vortex. \cite{ftmusr}
In Fig. 1 of Ref.\ \onlinecite{ftmusr} and the accompanying discussion,
it is shown that, for systems with very short coherence lengths,
the spatial dependence of the gap function outside the core has
a scale-independent power-law dependence, approaching its asymptotic
value roughly as $1/r^2$.  

The above discussion provides a natural explanation of the muon spin resonance results 
of Sonier and co-workers\cite{sonier} 
who found that the vortex core radius, defined as the radius at which 
the supercurrent density has its maximum, grows large at low
field.  In fact, an excellent fit to their data can be obtained by
assuming that the vortex core radius scales as $1/\sqrt{B}$, as is 
shown in Fig.\ \ref{fig:musr}. The coefficient of $1/\sqrt{B}$ from the fit is 
$r_0=46.3\pm 1.5 \AA T^{1/2}/\sqrt{B}$.  Since the vortex lattice constant 
$d$ for the $A$ or $B$ sublattices is $d=632 \AA T^{1/2}/\sqrt{B}$, this 
maximum occurs at about 7\% of $d$ or equivalently at about 10\% of
the intervortex spacing. It would be interesting to test this 
result at higher fields to see if this scaling breaks down and if $r_0$  
saturates at a constant value limited by the coherence
length $\xi_0$ as one might expect.

\section{The Density of States and the Specific heat}

\subsection{Semiclassical DOS}

We start by calculating the density of states for the semiclassical
(SC) approximation, in which the energy is Doppler shifted by the
local superfluid velocity $\bv_s(\br)$:
\begin{equation}
E(\bk,\br)=\hbar k_F \hat{x}\cdot\bv_s(\br)\pm\sqrt{(\hbar v_F k_x)^2+
(\hbar v_{\Delta} k_y)^2},
\end{equation}
where the spectrum has been linearized around the node $\vec k=(k_F, 0)$.
 
The local superfluid velocity far from the vortex is
$\bv_s(\br)=(\hbar/2mr)\hat{\phi}$.  In the commonly
employed ``single-vortex
approximation,'' the associated density of states is
\begin{eqnarray}
N(E)&=&2\frac{1}{\pi \ell^2}\int_{0}^\ell r dr d\varphi
\left\{\frac{V}{(2\pi)^2 w} \times \right. \\*
&{}&\left. \int_0^{E_c} \frac{2\pi\epsilon d\epsilon}{\hbar^2 v_F v_{\Delta}} 
\delta(E-\frac{\hbar^2 k_F}{2mr}\sin\varphi\mp\epsilon)\right\} \nonumber
\end{eqnarray}
where the factor of $2$ accounts for spin degeneracy.
$V/w$ is the total area of the CuO planes in the sample, where $V$ is the
volume of the sample and $w$ is the average separation between the planes,
and $\pi \ell^2=\Phi_0/B$ is the area of one unit cell of the vortex lattice. 
The integral is over $\epsilon=\sqrt{(\hbar v_F k_x)^2+
(\hbar v_{\Delta} k_y)^2}$.
   
In the absence of a magnetic field, with no Doppler shift,
\begin{equation} \label{eqn:N_0}
N_0(E)=\frac{V}{\pi\hbar^2 v_F \vd w}|E|.
\end{equation}
Putting the magnetic field back in, the density of states has the intercept
\begin{equation}
N(0)=\frac{2}{\pi}N_0\left(\frac{\hbar v_F}{\ell} \right) \label{eqn:vol_N(0)},
\end{equation}
where $N_0$ is the zero-field density of states. For nonzero $E$ we find that
\begin{equation} \label{eqn:vol_dos1}
N(\vep)=N(0)\left[\frac{
6\vep\sqrt{1-4\vep^2}+(8\vep^2+1)\sin^{-1}(2\vep)}{8\vep} \right] ,
\end{equation}
for $0\le|\vep| < 1/2$, where $\vep=E \ell/\hbar v_F$, and
\begin{equation} \label{eqn:vol_dos2}
N(\vep)=\frac{\pi}{2}N(0)\left(\vep+\frac{1}{8\vep}\right)
\end{equation}
for $|\vep|\ge1/2$.  Note that this is the contribution to the total density
of states for 2 spin states from \textit{one} of the four nodes.
 
A more realistic calculation of the semiclassical DOS can be made for a
square vortex lattice if we write the superfluid velocity as the Fourier sum
\begin{equation} \label{eqn:vs}
\bv_s(\br)=\frac{\pi\hbar}{m a^2}\sum_{\bQ\neq0}\frac{\im \bQ\times \hat{z}}
{Q^2}e^{i\bQ\cdot\br},
\end{equation}
where $\bQ=2\pi(m\hat{x}+n\hat{y})/a$ and $a=\sqrt{\Phi_0/B}$.
Note that
we are now orienting the {\emph{x}} and {\emph{y}} axes along the nearest-neighbor directions of 
the square vortex lattice.  The corresponding density of states, 
\begin{equation} N(E)=\frac{V/w}{\pi\hbar^2 v_F\vd a^2}\int_0^a\int_0^a dx dy |E-\hbar
\bv_s(\br)\cdot \bk_F|,
\end{equation}
can then be calculated numerically using this more accurate expression
for $\bv_s$.
The semiclassical density of states, as calculated for both the single-vortex
 approximation and the square vortex lattice, is shown in
Figs.\ \ref{fig:edos14} and \ref{fig:edos20}.  One can see that
the square-lattice DOS is about 30\% lower at zero energy than that calculated using the single-vortex approximation.  This lowering is caused by the disappearance of $\bv_s(\br)$ at high symmetry points on the vortex-lattice unit-cell boundary.\cite{vekhterp}

\subsection{Quantum DOS}
The quantum density of states is calculated from the quasiparticle energy spectrum at the node $\vec k=(k_F,0)$:
\begin{equation} \label{eqn:1ddos}
N(E)=2 \frac{V}{d^2 N_k w} \sum_{n\bk}\delta(E-E_{n\bk}),
\end{equation}
where n labels the energy bands and $\bk$ is a wave vector in the magnetic Brillouin zone. The factor of $2$ accounts for spin degeneracy. 
In order to clarify the dimensional analysis, we have multiplied 
the usual expression by 
$1=V/(w N_k d^2)$, where $N_k$ is the number of wave vectors in the MBZ, and $V/w$ 
is the total area of the CuO planes in the sample.  The energy in this expression is in 
units of $2\pi\hbar v_F/d$. In order to compare this result with the semiclassical 
result we simply write $N(E)$ in 
units of $N_0(\hbar v_F/\ell)$ [see Eq.\ \ref{eqn:N_0}], noting that 
$\ell=d/\sqrt{2\pi}$.  Results are shown in Figs.\ \ref{fig:edos14} and 
\ref{fig:edos20}, where comparison is made to both the 
SC single-vortex approximation and to the SC square-lattice DOS. Note that both axes scale as $1/\ell \propto \sqrt{B}$. 
The dotted line shows the commonly employed single vortex SC DOS to be roughly twice as large as the quantum 1-D DOS
in the low-energy region.  The quantum DOS rises more quickly with energy and
the SC and quantum DOS match up at higher energy andare
indistinquishable for energies above 3$E_V$.   The discrepancy between the SC
square-lattice calculation and the quantum DOS at low energies is due to
quantum effects that average over the rapid variations in the
direction of ${\bf v_s(r)}$ near the vortex cores as well as near the
high symmetry points on the unit-cell boundary.  Of course, disorder
effects on the vortex lattice and the quasiparticle energies will also
affect the average magnitude of the low-energy DOS in both the SC and
quantum cases.\cite{kubert,vekhterp} 

The 1-D calculation of $N(E)$ for $\ad=20$ (Fig.\ \ref{fig:edos20}) is in good 
agreement with the corresponding 2-D calculation of FT\cite{franz1}, reproducing all 
of the major features at low energies. The overall magnitude of the 1-D DOS is slightly reduced
from the full 2-D calculation. The 1-D calculation, by essentially averaging
in one direction, underestimates the
effect of the supercurrents, which push states to lower energies, as can be seen from
the bandstructures shown in Figs.\ \ref{fig:1d2d14} and \ref{fig:1d2d20}.  The full
2-D quantum DOS in Ref. \cite{franz1}, is about 10\% higher in magnitude than the
1-D approximation, but is still noticeably lower in magnitude than the SC
square-vortex-lattice result.

\subsection{Scaled $C_v(T,B)$}

The heat capacity of a fermion gas is
\begin{eqnarray}
C&=&2\beta k_B \sum_{\bk} -\frac{\partial f(E_k)}{\partial E_k} E_k^2 \nonumber \\*
&=&\frac{k_B}{\beta} \int_0^{\infty}N_T(u/\beta)\frac{u^2}{1+\cosh u}du\,. \label{eqn:c/t}
\end{eqnarray}
This is the expression used to calculate the specific heat ($C_v=C/V$) from the total density of states $N_T(E)$.  The total density of states in Eq.\ (\ref{eqn:c/t}) is a sum over the density of states \textit{for one spin} at each of the four nodes.  Thus, $N_T(E)=2\,N(E)$ where $N(E)$ is the semiclassical [Eqs.\ \ref{eqn:vol_dos1} \& \ref{eqn:vol_dos2}] or quantum mechanical [Eq.\ \ref{eqn:1ddos}] density of states calculated in the previous section.  Therefore, the specific heat at constant volume is
\begin{equation}
C_v=2 \frac{k_B^2 T}{V}\int_0^{\infty}N(u/\beta)\frac{u^2}{1+\cosh u}du.
\end{equation}
The specific heat for the 1-D calculation is shown for various values of $\ad$ 
in Fig.\ \ref{fig:covert}.  Again, both the $C/T$ axis and the $T$ axis scale 
as $\sqrt{B}$, in agreement with the general scaling predictions of 
Volovik\cite{volovik1}, and Simon and Lee\cite{simonlee}.  The $C_v/T$ is linear at 
higher temperatures, flattens out as the temperature is decreased, and then increases 
to a peak at even lower $T$ before rapidly falling, with a tiny shoulder on the way 
down (see inset of Fig.\ \ref{fig:covert}), to zero at $T=0$.  The large peak both 
sharpens and moves closer to the $T=0$ axis as the anisotropy $\ad$ is increased.   
The behavior of this peak suggests that its presence is due to the low energy
peaks in the DOS, particularly the van Hove singularities that occur just above
$E_V=0.25$ for $\ad=14$ as these contribute significant weight to
the DOS.   A narrow peak at $E$ in the DOS will typically show up as a peak in
the specific heat near $E/2$.\cite{kittel}  Comparing the 1-D and 2-D
dispersions and DOS, we expect this peak to shift to slightly lower energy and
to sharpen in the full 2-D calculation of the specific heat.  

The SC specific heat, for the square lattice and for the single-vortex approximation, 
is shown in Fig.\ \ref{fig:scaledcovert}, along with the 1-D specific heat.  The temperature is in units of $E_v=\hbar v_F/\ell$ and $C_v$ is in units of
\begin{equation}
\frac{k_B^2}{2\pi\hbar\vd w}\sqrt{\frac{\pi}{\Phi_0}}\sqrt{B}.
\end{equation}

Again, the main difference between the SC and quantum specific heat is that
the SC specific heat larger in magnitude.  Both exhibit the same
scaling with magnetic field and with $\alpha_D$.  The quantum specific heat
exhibits structure at the lowest temperatures which is a reflection of the
structure in the low-energy DOS. 

In order to make comparisons with experimental results, we use the numbers in 
Chiao \textit{et al.}\cite{chiao} for YBCO: $v_F\simeq2.5\times10^7 \textrm{cm/s}$, 
$\ad=14$, and $w=5.85$ {\AA}.  The molar volume of YBCO is  
$V_M=104.38~\textrm{cm}^3/\textrm{mole}$.\cite{simon}
With these numbers we obtain an intercept for the SC single vortex
calculation of $0.91\,(\textrm{mJ}\,\textrm{mol}^{-1})\textrm{K}^{-2} 
\textrm{T}^{-1/2}$ in seemingly excellent agreement with the experimental
$\sqrt{B}$ coefficient of $0.91\,(\textrm{mJ}\,\textrm{mol}^{-1})\textrm{K}^{-2}$ of
 Moler {\it et al.}\cite{kam}. However, since this approximation overestimates
the zero energy specific heat by roughly a factor of 2, this agreement is
fortuitous.   
The quantum specific heat for $\ad=14$ flattens out at 
approximately $0.5\,(\textrm{mJ}\,\textrm{mol}^{-1})\textrm{K}^{-2} \textrm{T}^{-1/2}$.

The Geneva group of Junod and co-workers has reported a number of 
results\cite{geneva,geneva2} for the specific heat of very high quality YBCO
crystals, grown in ${\rm BaZrO_3}$ crucibles and doped to ${\rm O_{7.00}}$
so as to minimize the effects of oxygen chain vacancies.  In the earlier of 
these, Revaz {\it et al.},\cite{geneva} the vortex contribution to the 
specific heat was obtained by subtracting $C(B\perp c,T)$ from 
$C(B\parallel c,T)$, the idea being that both lattice and magnetic impurity 
effects would cancel out in this subtraction and that the vortex contribution
to the spceific heat for $B\perp c$ is small. In the more recent preprint by
Wang {\it et al.}\cite{geneva2}, results are presented for $C(B,T)-C(0,T)$ 
for $B\parallel c$. For our purposes, these data are more directly useful since
they involve only the single field direction that we have studied.  Furthermore,
the results should be reliable, since the samples show very little sign of
point magnetic impurities.  Wang {\it et al.}\cite{geneva2} find a
$T\rightarrow 0$\ intercept for $(C(B,T)-C(0,T))/T$ of $1.34\pm 0.04$\ 
(mJ mol$^{-1}$)K$^{-2}$T$^{-1/2}$.

In order to compare our theoretical results to the $T$\ and $B$\ 
dependence found by Wang 
{\it et al.},\cite{geneva2} we need to subtract the specific heat in zero field
from that in a field.  The result is shown in the inset of Fig.\ 
\ref{fig:scaledcovert}. It is interesting that the structure
that we find at the lowest temperatures could easily be attributed
to Schottky-type anomalies in the data.  In fact Wang {\it et al.}
\cite{geneva2} show figures with and without subtraction of an assumed Schottky
anomoly, and the latter better resembles our theoretical results.  It is 
tempting to suggest that the experimentally observed low temperature structure
in samples with the least magnetic impurities is actually due to the structure
in the quasiparticle density of states.  However, since the magnitude of
the observed field-dependent specific heat is more than twice as large as the calculated
value, such detailed comparisons between theory and experiment are probably
premature.  The effect of disorder on the quasiparticles would likely
increase the low energy specific heat, since it increases the low energy DOS.
On the other hand, disorder in the vortex lattice may equally well decrease
the low-energy specific heat by reducing the local supercurrent velocity.
Therefore, it is not obvious that disorder, in itself, can account for
the differences between theory and experiment.  

\section{Conclusions}
By making the approximation, in the mixed state, that the low-energy 
quasiparticle states in the Dirac nodes are essentially one-dimensional, we 
have been able to obtain analytical results for the quasiparticle wave 
functions and energy spectra.  The 1-D approximation to the FT 
Hamiltonian\cite{franz1} elucidates the physics of the interaction of the 
quasiparticles in the lowest-energy bands with the vortex lattice:  the 
quasiparticles travel as plane-waves along the directions of the gap nodes, 
and are confined by the periodic potential of the vortex lattice in the 
direction of the Fermi surface.  Using these exact analytical results,
we were able to show that the approximate plane-wave solution for the same
problem converges rapidly.  The 1-D approximation is able to reproduce the 
important features of the 2-D plane-wave expansion in the lowest bands.

We have presented calculations of 
the tunneling density of states, which are in qualitative agreement with the 
semiclassical results of Mel'nikov\cite{melnikov2} but which also show spatial
structure due to the energy dispersion of the low-lying states.  
The density of states at zero
energy for the quantum problem is significantly lower (by a factor of 2) than
the commonly-employed semiclassical result for a single circular vortex,
although this simple approximation overestimates the semiclassical density of states 
for a square vortex lattice configuration 
by roughly 30\% at zero energy.
Thus this reduction arises from two sources:  the larger area of low
superfluid velocity in the Abrikosov lattice, compared to the case of a single
vortex in a circular unit cell of radius $\ell$, and quantum averaging of the superfluid velocity for
quasiparticles in the first magnetic Brillouin zone.

The specific heat has been calculated from the DOS of the 1-D plane-wave 
expansion and found to exhibit structure at low temperatures that is not present
in the semiclassical approximation.  In addition, the magnitude of the low-
temperature specific heat is reduced by quantum effects.  Since the values of the
specific heat measured experimentally,\cite{kam,geneva,geneva2} for parameters
$v_F$ and $v_\Delta$ taken from other experiments, are already
larger than the semiclassical results,  the disagreement in magnitude with the quantum
results are even larger.  One possibility  is that the discrepancy is due to
the effects of disorder which, to date, have not been included in any quantum
treatment of the specific heat.  Other possibilities are that the parameters
are such that $\ad$ is substantially larger than is currently believed, or that there are additional
low-energy states not accounted for in the disordered $d$-wave model but which exhibit similar magnetic-field dependence.

\acknowledgements  We would like to thank Marcel Franz, Peter Hirschfeld, 
Elisabeth Nicol, Zlatko Tesanovic, Ilya Vekhter and Rachel Wortis for useful 
discussions and correspondence and
Jeff Sonier for discussing and sharing his MuSR results with us.  We
thank the Institute for Theoretical Physics, where this work was completed,
for their hospitality.  This research was supported in part by the Natural
Sciences and Engineering Research Council (Canada) and by the National Science
Foundation under Grant No. PHY94-07194.

\appendix
\section{Analytical solution of the 1-D problem}
\label{1dappendix}
With the approximation that the potential is one-dimensional the quasiparticle Hamiltonian is
\begin{equation}
{\mathcal H}_{1\text{D}}=\left( 
\begin{array}{cc}
q_x^n   &  -\frac{\im}{2 \pi \ad} \frac{d}{dz} \\
 -\frac{\im}{2 \pi \ad} \frac{d}{dz} & -q_x^n
\end{array}
\right)
+
\left( 
\begin{array}{cc}
\Phi(z-\frac{1}{4})  &  0 \\
0  &  \Phi(z+\frac{1}{4})
\end{array}
\right).
\end{equation}
The Hamiltonian can be rewritten as
\begin{equation}
\label{eqn:H2}
{\mathcal H}_{1\text{D}}=\hat{\sigma}_z q_x^n + \hat{\sigma}_x \left( -\frac{\im}{2 \pi\ad}\frac{d}{dz} \right)+\frac{\Phi_1(z)}{2}+\hat{\sigma}_z\frac{\Phi_2(z)}{2},
\end{equation}
where $\Phi_1(z)=\Phi(z-\frac{1}{4})+\Phi(z+\frac{1}{4})$, and $\Phi_2(z)=\Phi(z-\frac{1}{4})-\Phi(z+\frac{1}{4})$. Borrowing a trick from Mel'nikov   \cite{melnikov1}, one can insert ${\bf I}=1/2 (\hat{\sigma}_x+\hat{\sigma}_z)(\hat{\sigma}_x+\hat{\sigma}_z)$ between ${\mathcal{H}}_{1\rm{D}}$ and $\psi$ and then multiply Eq.\ (\ref{eqn:H2}) on the left by $1/2(\hat{\sigma}_x+\hat{\sigma}_z)$.  This transformation takes $\hat{\sigma}_x \to \hat{\sigma}_z$ and vice versa.  We then write 
\begin{displaymath}
\varphi(z) = 1/2(\hat{\sigma}_x+\hat{\sigma}_z)\psi(z),
\end{displaymath}
so that 
\begin{eqnarray}
\left\{\hat{\sigma}_z \left(-\frac{\im}{a}\frac{d}{dz}\right)+\frac{\Phi_1(z)}{2}-\epsilon+\hat{\sigma}_x\left[q_x^n+\frac{\Phi_2(z)}{2}\right] \right\} \varphi(z) \nonumber \\*
=0,
\end{eqnarray} 
where $a=2\pi\ad$.  Writing $\varphi(z)=(f(z),g(z))^T$ we obtain the following coupled first-order differential equations:
\begin{eqnarray*} 
\left(-\frac{\im}{a}\frac{d}{dz}+\frac{\Phi_1(z)}{2}-\epsilon \right)f(z)+\left(q_x^n+\frac{\Phi_2(z)}{2}\right)g(z) &=& 0 \\*
\left(q_x^n+\frac{\Phi_2(z)}{2}\right)f(z)+\left(\frac{\im}{a}\frac{d}{dz}+\frac{\Phi_1(z)}{2}-\epsilon \right)g(z) &=& 0. 
\end{eqnarray*}
From these coupled equations we can derive a second order differential equation for $f(z)$:
\begin{eqnarray}
f''(z)&+&a^2\left[\frac{\im}{a}+\left(\frac{\Phi_1(z)}{2}-\epsilon\right)^2-\left(q_x^n+\frac{\Phi_2(z)}{2}\right)^2\right]f(z) \nonumber \\*
&=& \frac{\im a}{2}(f(z)-g(z))\delta((z-n) \label{eqn:deltaDQ} + \frac{1}{4})  \\*
&+&\frac{\im a}{2}(f(z)+g(z))\delta((z-n)-\frac{1}{4}) \nonumber
\end{eqnarray}
with delta functions at the boundaries and at the center of the unit cell.  In the regions $-\frac{1}{4}<z-n<\frac{1}{4}$ and $\frac{1}{4}<z-n<\frac{3}{4}$, the second order differential equation for $f(z)$ is
\begin{eqnarray} 
f''(z)&+&a^2 \left[\frac{\im}{a}+\left(\frac{\Phi_1(z)}{2}-\epsilon \right)^2-\left(q_x^n+\frac{\Phi_2(z)}{2}\right)^2\right]f(z)\nonumber \\*
&=&0. 
\label{eqn:DQ} 
\end{eqnarray}
Since this differential equation is periodic in $z$, we can solve it within a unit cell and use Bloch's theorem to extend the solution over all of $z$.  Since $\Phi_1(z)$ has a period of $1/2$ and  $\Phi_2(z)$ has a period of $1$, we divide the unit cell into two regions (taking $n=0$ for simplicity): $-\frac{1}{4}<z<\frac{1}{4}$ and $\frac{1}{4}<z<\frac{3}{4}$.  In these two regions
\begin{displaymath}
\Phi_1(z)=
\left\{ 
\begin{array}{r@{,~}r@{\:<\:}c@{\:<\:}l}
2 z&-1/4&z&1/4 \\
2 z-1&1/4&z&3/4
\end{array} \right.
\end{displaymath}
and
\begin{displaymath}
\Phi_2(z)=
\left\{ 
\begin{array}{r@{,~}r@{\:<\:}c@{\:<\:}l}
+1/2&-1/4&z&1/4\\
-1/2&1/4&z&3/4
\end{array} \right. .
\end{displaymath}
Taking region 1 as $-\frac{1}{4}<z<\frac{1}{4}$ and region 2 as $\frac{1}{4}<z<\frac{3}{4}$ we can write
\begin{equation}
f''(z)+a^2\left((z-c_j)^2-b_j^2+\frac{\im}{a}\right)f(z)=0,
\end{equation}
where $j=1,2$ and $b_1=q_x^n+\frac{1}{4}$, $b_2=q_x^n-\frac{1}{4}$, $c_1=\epsilon$ and $c_2=\epsilon+\frac{1}{2}$.  If we let $\tau_j=\sqrt{a}(z-c_j)$ then
\begin{equation}
f''(\tau_j)+\left(\tau_j^2-a b_j^2+\im \right)f(\tau_j) = 0.
\end{equation}
This equation is solved by the parabolic cylinder functions (see Gradshteyn and Ryzhik \cite{g&r})
\begin{equation}
f(\tau_j)=D_{\im\lambda_j}\left[\pm(1+\im)\tau_j \right],
\end{equation}
where $\lambda_j=a b_j^2/2$.  The corresponding solution for $g(z)$ is easily obtained. We thus obtain the full solution for $\varphi(z)$ shown in Eqs.\ (\ref{eqn:solution1}) and (\ref{eqn:solution2}).

\section{Solution of the boundary conditions to obtain an excitation spectrum}
\label{bcappendix}
For convenience we rewrite Eqs.\ (\ref{eqn:solution1}) and (\ref{eqn:solution2}) as
\begin{equation}
\varphi_1(z)=\left(
\begin{array}{c}
A_n~f_1^+(z) + B_n~f_1^-(z) \\ 
-A_n~g_1^+(z) + B_n~g_1^-(z)
\end{array} \right),
\end{equation}
for $n-\frac{1}{4}<z<n+\frac{1}{4}$ and
\begin{equation}
\varphi_2(z)=\left(
\begin{array}{c}
C_n~f_2^+(z) + D_n~f_2^-(z) \\ 
-C_n~g_2^+(z) + D_n~g_2^-(z)
\end{array} \right),
\end{equation}
for $n+\frac{1}{4}<z<n+\frac{3}{4}$.

Acceptable solutions must be continuous at the interior point, $z=n+\frac{1}{4}$, and at the boundaries of the unit cell, $z=n-\frac{1}{4}$.  At $z=n+\frac{1}{4}$ the boundary condition is
\begin{eqnarray*}
\left( \begin{array}{c}
A_n~f_1^+(n+\frac{1}{4}) + B_n~f_1^-(n+\frac{1}{4}) \\ 
-A_n~g_1^+(n+\frac{1}{4}) + B_n~g_1^-(n+\frac{1}{4})
\end{array} \right) ~~~~~~~~~~~~~~~~~~\\*
~~~~~~~~~~~~=\left( \begin{array}{c}
C_n~f_2^+(n+\frac{1}{4}) + D_n~f_2^-(n+\frac{1}{4}) \\ 
-C_n~g_2^+(n+\frac{1}{4}) + D_n~g_2^-(n+\frac{1}{4})
\end{array} \right), 
\end{eqnarray*}
which can be rewritten as
\begin{equation}
{\bf M_1} 
\left(\begin{array}{c}
A_n \\
B_n
\end{array}\right)
=
{\bf M_2} 
\left(\begin{array}{c}
C_n \\ 
D_n
\end{array}\right). 
\label{n}
\end{equation}
At $z=n-\frac{1}{4}$ the boundary condition is 
\begin{eqnarray*}
\left( \begin{array}{rc}
f_1^+(n-\frac{1}{4})&f_1^-(n-\frac{1}{4}) \\[0.1cm] 
-g_1^+(n-\frac{1}{4})&~g_1^-(n-\frac{1}{4})
\end{array} \right)
\left(\begin{array}{c}
A_n \\ 
B_n
\end{array}\right) ~~~~~~~~~~~~~~~~~~~~~~~~~~ \\*
~~~~~~~~~~~~~=\left( \begin{array}{rc}
f_2^+(n+\frac{3}{4})&f_2^-(n+\frac{3}{4}) \\[0.1cm] 
-g_2^+(n+\frac{3}{4})&g_2^-(n+\frac{3}{4})
\end{array} \right)
\left(\begin{array}{c}
C_{n-1} \\
D_{n-1}
\end{array}\right),
\end{eqnarray*}
or
\begin{equation}
{\bf M_3} 
\left(\begin{array}{c}
A_n \\
B_n
\end{array}\right)
=
{\bf M_4} 
\left(\begin{array}{c}
C_{n-1} \\
D_{n-1}
\end{array}\right),
\label{n-1} 
\end{equation}
where we have used the periodicity of $\varphi_2(z)$ on the right hand side of the above equation.  From Eqs.\ (\ref{n}) and (\ref{n-1}) we can write 
\begin{equation}
\left(\begin{array}{c}
C_{n} \\
D_{n} 
\end{array} \right)
={\bf P}\left(\begin{array}{c}
C_{n-1} \\
D_{n-1} 
\end{array} \right),
\end{equation}
where
\begin{equation}
{\bf P}={\bf M_2}^{-1}{\bf M_1}~{\bf M_3}^{-1}{\bf M_4}.
\end{equation}
Since the Hamiltonian is periodic the eigenvalues $p$ of ${\bf P}$, an operator which induces a translation of one period, must satisfy the Bloch condition:
\begin{equation}
p_{\pm}=e^{\pm 2\pi\im q_y},
\end{equation}
where $-\frac{1}{2}<q_y<\frac{1}{2}$.
The eigenvalues of ${\bf P}$ are the roots of the characteristic equation
\begin{equation}
p^2-p~tr({\bf P})+|{\bf P}| = 0.
\end{equation}
Clearly
\begin{equation}
p_{\pm}=\frac{1}{2} \left(tr({\bf P}) \pm \sqrt{(tr({\bf P}))^2-4}\right),
\end{equation}
which implies that 
\begin{equation}
\frac{p_{+} +p_{-}}{2}=\cos(2\pi q_y)=\frac{1}{2}~ tr({\bf P}).
\end{equation}
Since $q_y$ is real 
\begin{equation}
\text{Im}\{tr({\bf P})\}=0
\end{equation}
and
\begin{equation}
\label{eqn:realtrace}
\text{Re}\{tr({\bf P})\}=2\cos(2\pi q_y).
\end{equation}
The energy spectrum can now be directly calculated using this
expression.

\pagebreak

\begin{center}	
\begin{figure}  
\scalebox{0.7}{\includegraphics*{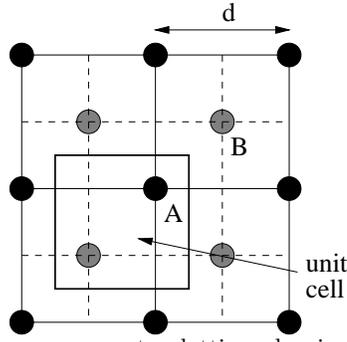}}
\caption{The square-vortex lattice, showing the $A$ and $B$ sublattices and 
the corresponding unit cell.  The edges of the unit cell are aligned with
the {\emph{x}} and {\emph{y}} axes that are the nodal directions.} 
\label{fig:lattice}     
\end{figure}    

\begin{figure}
\scalebox{0.6}{\includegraphics*{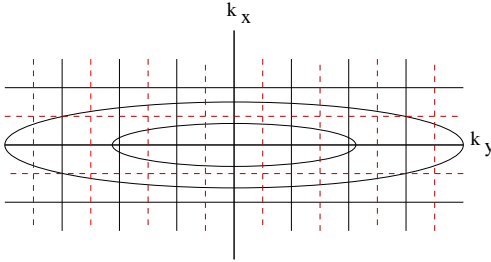}}
\caption{Constant-energy contours of $\cH_0$ and MBZ boundaries of the A and B square sublattices at $\ad\simeq 5$.}  
\label{fig:mbz}
\end{figure}    

\begin{figure}
\scalebox{0.48}{\includegraphics*{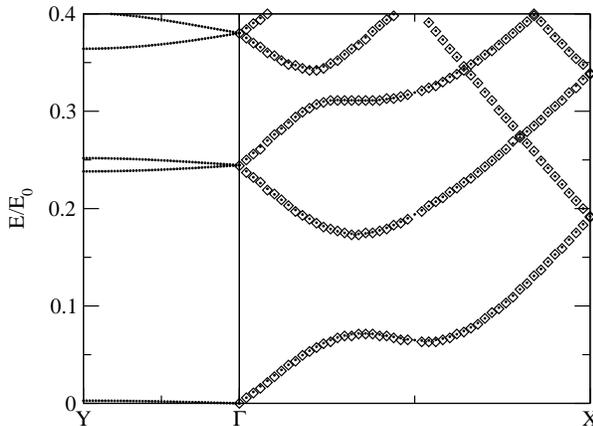}}   
\caption{A comparison of the 1-D analytical spectrum along the $k_x$ axis 
($\diamond$) with the numerical 1-D plane-wave expansion (61 RLVs) results ($\cdot$) for $\ad=7$.  The numerical 1-D results are also shown for the spectrum along the $k_y$ axis.}
\label{fig:anal7}
\end{figure}    

\begin{figure}
\scalebox{0.48}{\includegraphics*{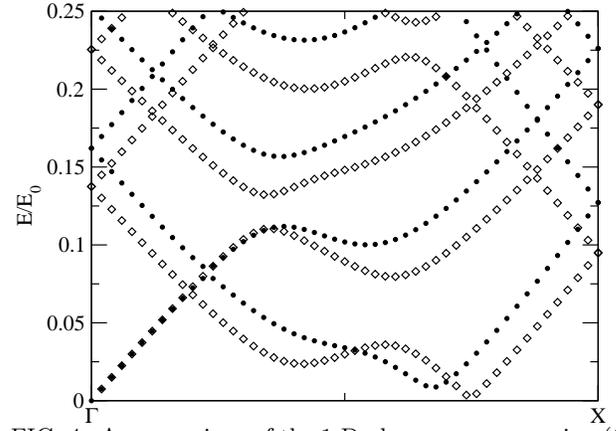}}   
\caption{A comparison of the 1-D plane-wave expansion (61 RLVs) ($\circ$) 
with the 2-D plane-wave expansion ($33\times$33 RLVs) results ($\diamond$) 
for $\ad=14$}
\label{fig:1d2d14}
\end{figure}    

\begin{figure}
\scalebox{0.48}{\includegraphics*{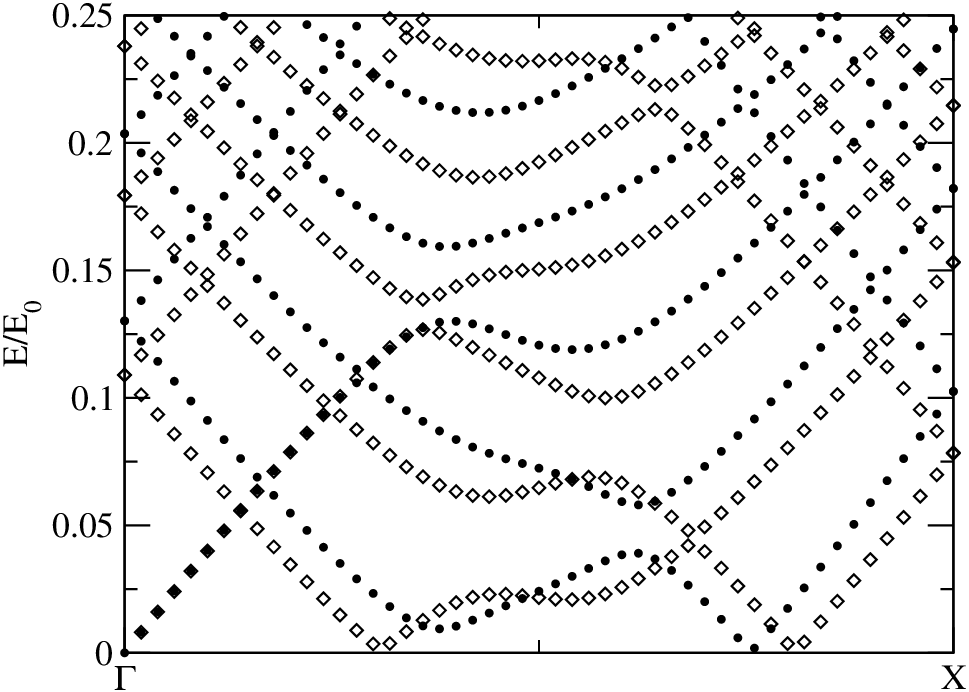}}
\caption{As in Fig.\ \ref{fig:1d2d14} for $\ad=20$}
\label{fig:1d2d20}
\end{figure}    

\begin{figure}
\scalebox{0.48}{\includegraphics*{fig6.eps}}   
\caption{The energy spectrum for $\ad =14$ and 
$N_y=41$ and $N_x=1 (\triangle),\ 5 (\times),\ 9 (\bullet),\ 13
(\ast),\ 21 ({\vrule height .9ex width .8 ex depth -.1ex}),\ {\rm and}\ 29 (\circ).$ }
\label{fig:conv14}      
\end{figure}    

\begin{figure}
\scalebox{0.48}{\includegraphics*{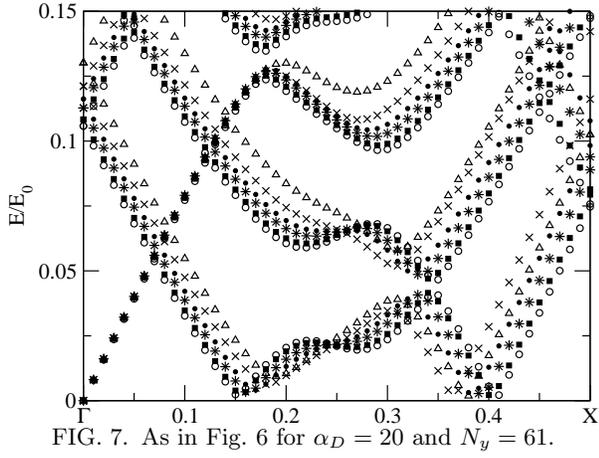}}   
\caption{As in Fig.\ \ref{fig:conv14} for $\ad=20$ and $N_y=61$.}
\label{fig:conv20}
\end{figure}    
		
\begin{figure}
\scalebox{0.48}{\includegraphics*{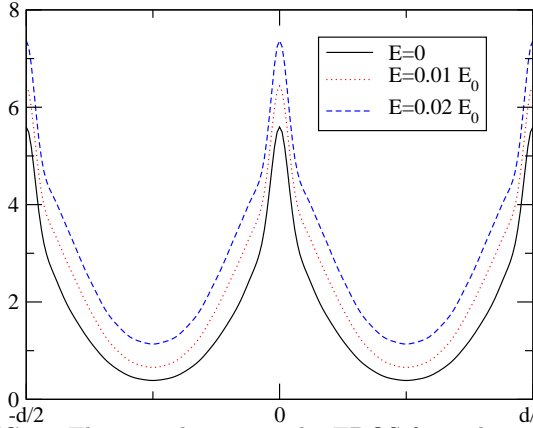}}  	
\caption{The contribution to the TDOS from the nodes at $\vec k=(\pm k_F,0)$ at three different energies for $\ad=14$. The TDOS is normalized as in Eq.\ (\ref{eqn:ldos}). 
$d/2$ is the separation, in the {\emph{y}} direction, between lines of vortices.  Note the shoulders forming on either side of the peaks as the energy is increased.}
\label{fig:ldos14}      
\end{figure}    

\begin{figure}
\scalebox{0.48}{\includegraphics*{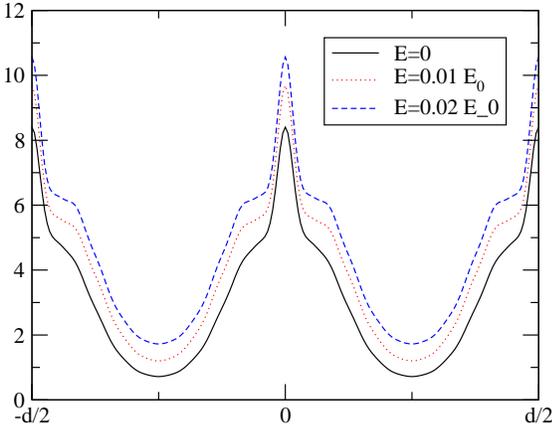}}   
\caption{As in Fig.\ \ref{fig:ldos14} for $\ad=20$.  Note the now very distinct shoulders which have formed on either side of the peaks.}
\label{fig:ldos20}      
\end{figure}    

\begin{figure}
\scalebox{0.8}{\includegraphics*{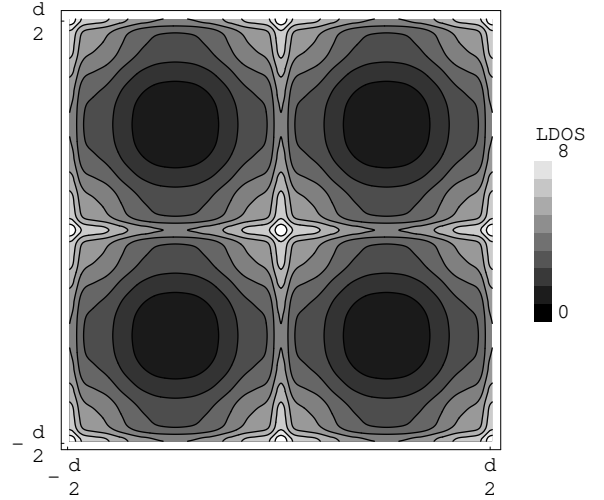}}   
\caption{The zero bias TDOS for the 1-D plane-wave expansion at $\ad=20$}
\label{fig:ldos2d}      
\end{figure}

\begin{figure}
\scalebox{0.48}{\includegraphics*{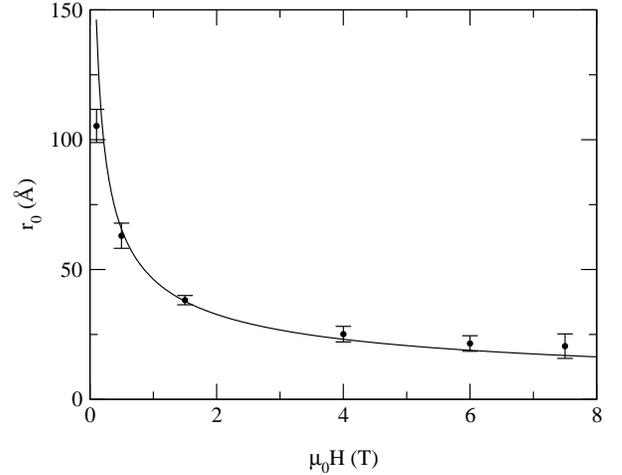}}          
\caption{A fit of the magnetic field dependence of the vortex-core radius 
as determined from muon spin resonance 
to the $1/\sqrt{B}$ (Ref.\ 25) scaling expected from our analysis.}            
\label{fig:musr}                
\end{figure}            

\begin{figure}
\scalebox{0.48}{\includegraphics*{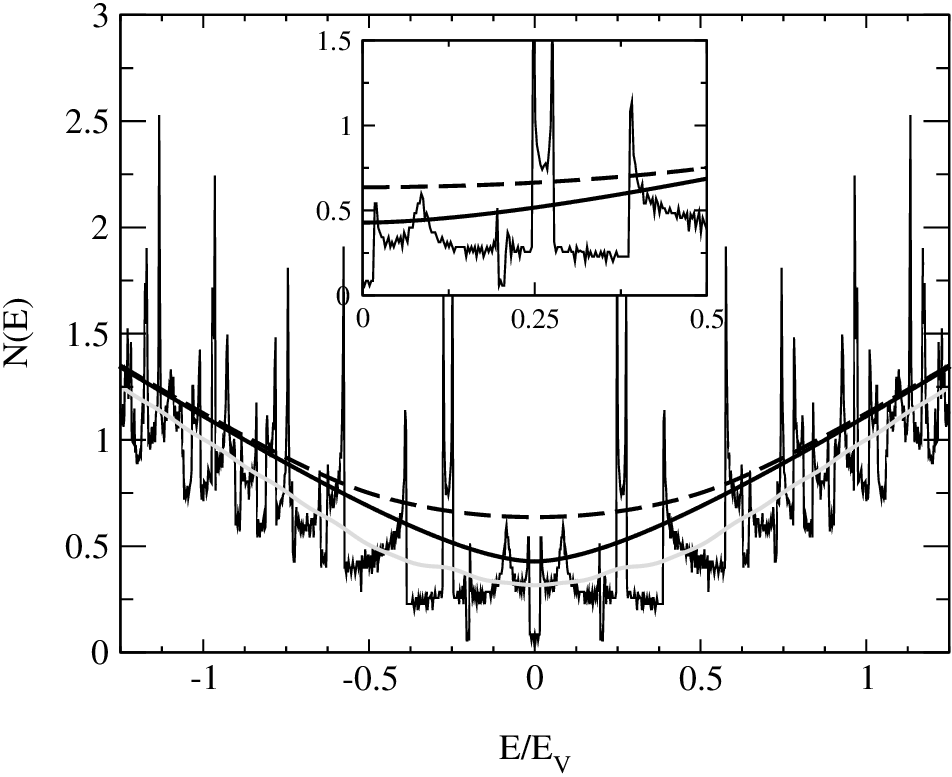}}  
\caption{The total DOS in units of $N_0(\hbar v_F/\ell)$ for the node 
$\vec k=(k_F, 0)$ and $\ad=14$, scaled to show the correspondence with the 
SC  calculation for the square-vortex lattice (solid line) and in the single-vortex
approximation (dashed line).  Note that both axes scale as $\sqrt{B}$. 
The energy is in units of $E_v=\hbar v_F/\ell$.  Also shown (thick solid line) 
is the ``averaged'' quantum DOS, broadened with a Gaussian of full width 0.08 
$E_v$.  The inset shows the low energy DOS compared to the SC approximations.
The averaged quantum DOS is not shown in the inset.}
\label{fig:edos14}
\end{figure} 

\begin{figure}
\scalebox{0.48}{\includegraphics*{fig13.eps}}
\caption{As in Fig.\ \ref{fig:edos14} for $\ad=20$.}
\label{fig:edos20}
\end{figure}

\begin{figure}
\scalebox{0.48}{\includegraphics*{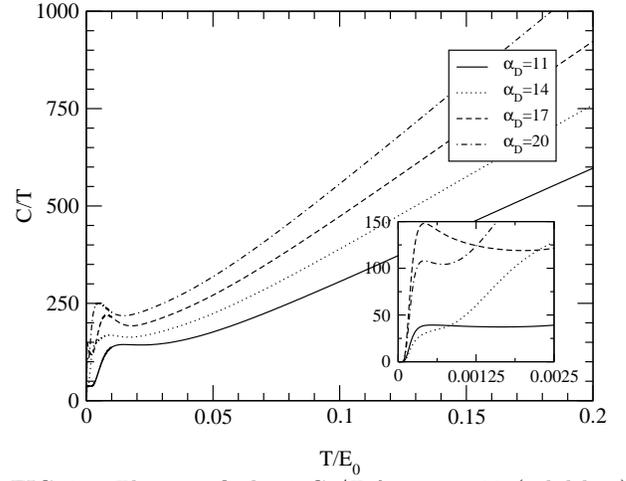}}
\caption{The specific heat $C_v/T$ for $\ad=11$ (solid line), $\ad=14$ (dotted line), $\ad=17$ (dashed line), and $\ad=20$ (dash-dot line).  The inset shows a magnification near $T=0$ of the same.}
\label{fig:covert}
\end{figure}

\begin{figure}
\scalebox{0.48}{\includegraphics*{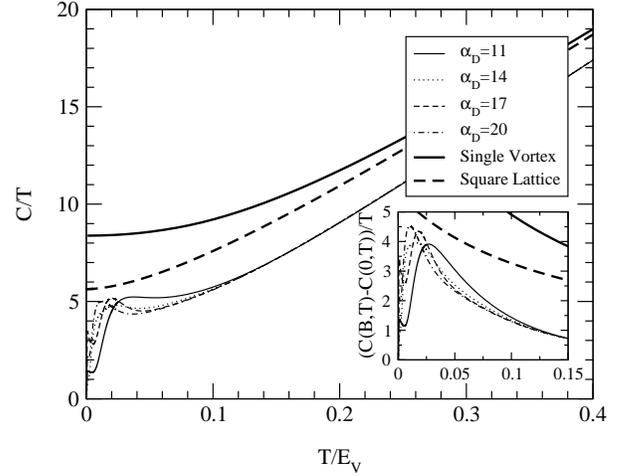}}
\caption{The specific heat, scaled as $1/\ad$ to show the correspondence with 
$C_v/T$ calculated from the Doppler-shifted energy spectrum.  The inset shows
the specific heat with the zero magnetic field value subtracted as is done
in Ref. 26.}
\label{fig:scaledcovert}
\end{figure}
\end{center}

}\end{multicols}

\begin{references}
\bibitem{convergence} In the 1-D plane-wave expansions, convergence 
is achieved by keeping 41 RLVs for $\alpha_D=14$ and 61 RLVs
for $\alpha_D=20$.
\end{references}
\end{document}